\documentclass[12pt]{article}
\usepackage{epsfig}
\usepackage{graphicx}

\def\beq{\begin{equation}}
\def\eeq#1{\label{#1}\end{equation}}
\def\eeqn{\end{equation}}

\def\beqa{\begin{eqnarray}}
\def\eeqa#1{\label{#1}\end{eqnarray}}
\def\eeqan{\end{eqnarray}}

\let\bar=\overbar

\def\Dslash{\not{\hbox{\kern-4pt $D$}}}
\def\dslash{\not{\hbox{\kern-2pt $\del$}}}

\def\msb{{\bar{\ssstyle M \kern -1pt S}}}

\usepackage{fancyhdr,graphicx}
\def\Title#1{\begin{center} {\Large {\bf #1} } \end{center}}

\begin{document}
\newcommand{\bemy}{\begin{equation}}
\newcommand{\eemy}{\end{equation}}
\newcommand{\bea}{\begin{eqnarray}}
\newcommand{\eea}{\end{eqnarray}}
\newcommand{\ba}{\begin{array}}
\newcommand{\ea}{\end{array}}
\newcommand{\bc}{\begin{center}}
\newcommand{\ec}{\end{center}}

\Title{Phase Structure and Transport Properties\\[0.2cm] 
of Dense Quark Matter}

\bigskip\bigskip


\begin{raggedright}
{\it Thomas Sch\"afer\\
Department of Physics\\
North Carolina State University\\
Raleigh, NC 27695}
\bigskip\bigskip
\end{raggedright}


\section{Introduction}     
\label{sec_intro} 
 
  This goal of this meeting is the search for exotic states of 
matter in compact astrophysical objects. Over the years many 
possible signatures for phase transitions in compact stars have 
been suggested, for example unusual masses, radii or cooling 
histories, or sudden changes in the spin frequency. In this 
contribution we will take a very conservative approach and 
study the predictions of weak coupling QCD for the equilibrium 
and transport properties of the densest phases of QCD. This 
includes the color-flavor-locked (CFL) phase \cite{Alford:1999mk}, 
and phases that arise from modifications of the basic CFL pairing 
pattern due to the effects of the non-zero strange quark mass.
A flavor rotation of the CFL condensate leads to kaon 
condensation (CFL-K) \cite{Bedaque:2001je}, and a spatial 
modulation of the CFL state leads to the meson supercurrent 
state (curCFL) \cite{Schafer:2005ym,Kryjevski:2005qq}. 

 We refer the reader to our recent review \cite{Alford:2007xm}
for a detailed discussion of quark matter phases at lower 
density. Some of these phases are stable in weak coupling, and 
their properties are rigorously computable in perturbative
QCD. This includes crystalline color superconductivity 
\cite{Alford:2000ze,Casalbuoni:2003wh}, and single flavor 
spin-one color superconductivity \cite{Schafer:2000tw,Schmitt:2004et}.
Other phases require strong coupling, for example the 2SC phase 
\cite{Rapp:1997zu,Alford:1997zt}, chiral density waves (and the 
quarkyonic phase) \cite{Deryagin:1992rw,Shuster:1999tn,Kojo:2009ha}, 
or gapless color superconductivity \cite{Shovkovy:2003uu,Alford:2003fq}.

 The goal of our research program is to compute the properties
of all these phases, and to determine whether hybrid neutron/quark 
matter stars containing quark matter in the CFL phase or one of 
the lower density quark phases are consistent with observation.
Current data on the masses and radii of compact stars are 
consistent with pure neutron stars as well as with hybrid quark 
matter stars \cite{Alford:2004pf,Kurkela:2009gj}. There are some 
results that indicate that other observational properties also do not 
clearly distinguish neutron stars from hybrid quark stars. In that 
case evidence for the existence of a high density phase in compact 
stars may have to come from careful studies of the mass-radius 
relation of compact stars \cite{Lattimer:2006xb}, or from the 
observation of gravitational waves from binary compact star mergers. 
Laboratory experiments involving heavy ions have established 
meaningful constraints on the equation of state of dense matter 
\cite{Danielewicz:2002pu}, and the next generation of these 
experiments may demonstrate the existence of a first order 
transition at densities that are achieved in compact stars. 

\section{Matter at the highest densities}     
\label{sec_phase}
\subsection{The CFL Phase}
\label{sec_cfl} 
 
  Calculations based on weak-coupling QCD indicate that the 
ground state of three flavor baryonic matter at the highest 
densities is the color-flavor-locked (CFL) phase 
\cite{Alford:1999mk,Schafer:1999fe,Evans:1999at}. The CFL phase
is characterized by a pair condensate
\bemy
\label{cfl}
 \langle \psi^a_i C\gamma_5 \psi^b_j\rangle  = 
  (\delta^a_i\delta^b_j-\delta^a_j\delta^b_i) \phi .
\eemy
This condensate leads to a gap in the excitation spectrum 
of all fermions and completely screens the gluonic interaction. 
Both the chiral $SU(3)_L\times SU(3)_R$ and color $SU(3)$ 
symmetry are broken, but a vector-like $SU(3)$ flavor symmetry 
remains unbroken. 

 The gap in the fermion spectrum can be computed in perturbative 
QCD. The nine quarks species (three flavors and three colors)
form an octet and a singlet under the unbroken $SU(3)$ flavor 
symmetry. The octet and singlet quarks have gaps $\Delta_0$ and 
$2\Delta_0$, respectively, where 
\cite{Son:1999uk,Brown:1999aq,Wang:2001aq,Schafer:2003jn}
\bemy
\label{gap_oge}
\Delta_0 \simeq 2^{-1/3} 512\pi^4  \mu
   \left(\frac{2}{3g^2}\right)^{5/2} 
   \exp\left(-\frac{3\pi^2}{\sqrt{2}g}-\frac{\pi^2+4}{8}\right)\, .
\eemy
Here, $\mu$ is the chemical potential and $g$ is the strong coupling 
constant. At this order in $g$ we can compute the coupling by 
evaluating the one-loop running coupling at the scale $\mu$. At large 
$\mu$ the coupling constant is small and the gap is exponentially small 
compared to the Fermi energy $E_f=\mu$. For densities relevant to 
neutron stars $\mu<500$ MeV and the coupling is not small. In this 
regime higher order correction to equ.~(\ref{gap_oge}) are not small
and the gap is quite uncertain. It is remarkable, however, that 
both extrapolations of the weak coupling result to the regime of 
moderate coupling \cite{Schafer:1999jg} as well as model calculations 
based on Nambu-Jona Lasinio (NJL) or similar interactions 
\cite{Buballa:2003qv} give gaps in the range $\Delta_0 \simeq 
(50-100)$ MeV at a density $\rho\simeq 5\rho_0$, where $\rho_0$ 
is the nuclear matter saturation density. 

 Perturbation theory can also be used to compute the gluon screening 
mass. Screening arises from the Meissner effect, as it does in 
ordinary superconductors. We find \cite{Son:1999cm,Rischke:2000ra}
\bemy 
 m^2 = \frac{21-8\log(2)}{54}\frac{g^2\mu^2}{2\pi^2}\, , 
\eemy
which shows that the gluon screening length is short compared to 
the coherence length $\xi\sim \Delta_0^{-1}$. We conclude that color 
superconductivity is type I. We note that electromagnetism
is unbroken, and the photon is not screened. The physical photon
in the CFL phase is a mixture of the bare photon and the bare gluon, 
and interesting effects occur at the interface of normal nuclear 
matter and the CFL phase \cite{Manuel:2001mx}.

\subsection{Effective theory of the CFL phase}
\label{sec_CFLchi}

 For excitation energies smaller than the gap the only relevant 
degrees of freedom are the Goldstone modes associated with the 
breaking of $SU(3)_L\times SU(3)_R$ chiral symmetry, the $U(1)_B$ 
phase symmetry associated with baryon number, and the approximate
$U(1)_A$ phase symmetry. The interaction of the low energy 
modes is described by an effective field theory. The structure 
of the effective lagrangian is determined by the symmetries of the 
CFL phase, and the coefficients that appear in the lagrangian 
can be computed in perturbative QCD. We will see that, with one
notable exception, these coefficients agree with estimates
based on naive dimensional analysis. 

 The effective lagrangian for the low energy modes has two 
important applications. First, it determines the spectrum and 
the interactions of quasi-particles. Based on this knowledge 
we can compute the specific heat and the transport properties 
of the CFL phase. Second, the effective theory determines 
the response of the CFL phase to non-zero quark masses and 
to external fields, such as lepton chemical potentials or magnetic 
fields. The effective lagrangian therefore determines the phase
structure at non-asymptotic densities. 

 The Goldstone mode associated with superfluidity is related
to the phase $\varphi$ of the order parameter 
\bemy 
\label{order_ph}
\langle\psi^\alpha_i C\gamma_5\psi^\beta_j\rangle
  = \epsilon^{\alpha\beta A}\epsilon_{ij B} 
     e^{2i\varphi}\phi_A^B \, , 
\eemy
where $\phi_A^B$ parametrizes the color-flavor orientation of 
the order parameter. The field $\varphi$ transforms as $\varphi\to
\varphi+\alpha$ under $U(1)_B$ transformation of the quark fields 
$\psi\to \exp(i\alpha)\psi$. At leading order in the weak coupling 
limit the effective lagrangian is completely fixed by Lorentz 
invariance and $U(1)_B$ symmetry. One can show that \cite{Son:2002zn}
\bemy 
\label{u1_eft_2}
{\cal L} = \frac{3}{4\pi^2}\left[ (\partial_0\varphi-\mu)^2 -
  (\nabla\varphi)^2 \right]^2 + \ldots ,
\eemy
where $\ldots$ denotes terms that are higher order in $g$, or 
terms of the form $\partial^n\varphi^m$ with $n>m$. In the low 
energy limit we can expand the lagrangian in powers of $\partial
\varphi$. We also rescale the field as $\phi=(3\mu/\pi)\varphi$ 
in order to make it canonically normalized. We will refer to 
$\phi$ as the phonon field. The phonon lagrangian is 
\bemy
\label{u1_eft_3}
{\cal L} = \frac{1}{2} (\partial_0 \phi)^2 - 
  \frac{1}{2} v^2 (\partial_i \phi)^2 - 
  \frac{\pi}{9\mu^2} \partial_0 \phi(\partial_\mu \phi \partial^\mu \phi) 
+ \frac{\pi^2}{108\mu^4}(\partial_\mu \phi \partial^\mu \phi)^2 
+ \ldots \, , 
\eemy
where $v=1/\sqrt{3}$ is the speed of sound. We observe that three and 
four phonon vertices are suppressed by powers of $|\partial\phi|/\mu^2$. 

 The effective lagrangian for the Goldstone modes associated with 
chiral symmetry breaking has the same structure as the chiral 
lagrangian at $T=\mu=0$. The main difference is that Lorentz invariance 
is broken and only rotational invariance is a good symmetry. The 
effective lagrangian is given by \cite{Casalbuoni:1999wu}
\bea
\label{l_cheft}
{\mathcal L}_{eff} &=& \frac{f_\pi^2}{4} {\rm Tr}\left[
 \nabla_0\Sigma\nabla_0\Sigma^\dagger - v_\pi^2
 \partial_i\Sigma\partial_i\Sigma^\dagger \right] 
 +\left[ B {\rm Tr}(M\Sigma^\dagger) + h.c. \right] 
    \nonumber \\ 
 & & \hspace*{0cm}\mbox{} 
     +\left[ A_1{\rm Tr}(M\Sigma^\dagger)
                        {\rm Tr} (M\Sigma^\dagger) 
     + A_2{\rm Tr}(M\Sigma^\dagger M\Sigma^\dagger) \right.
 \nonumber \\[0.1cm] 
  & &   \hspace*{0.5cm}\mbox{}\left. 
     + A_3{\rm Tr}(M\Sigma^\dagger){\rm Tr} (M^\dagger\Sigma)
         + h.c. \right]+\ldots . 
\eea
Here $\Sigma=\exp(i\phi^a\lambda^a/f_\pi)$ is the chiral field,
$f_\pi$ is the pion decay constant and $M$ is the mass matrix.
The chiral field and the mass matrix transform as $\Sigma\to 
L\Sigma R^\dagger$ and  $M\to LMR^\dagger$ under chiral 
transformations $(L,R)\in SU(3)_L\times SU(3)_R$. In order 
to determine the structure of the effective theory we treat 
$M$ as a field, but in practice we are interested in the case
$M={\rm diag}(m_u,m_d,m_s)$.

 The coefficients $f_\pi,v_\pi$, $B,A_i,\ldots$ can be computed 
in weak coupling perturbation theory. In the case of $f_\pi$ and
$v_\pi$ this is most easily done my matching the screening masses
for flavored gauged fields. The coefficients $B,A_i$ are related 
to the quark mass dependence of the condensation energy in the 
CFL phase. 

 At leading order in $\alpha_s$ the Goldstone boson decay 
constant and velocity are \cite{Son:1999cm} 
\bemy
\label{cfl_fpi}
f_\pi^2 = \frac{21-8\log(2)}{18} 
  \left(\frac{p_F^2}{2\pi^2} \right), 
\hspace{0.5cm} v_\pi^2=\frac{1}{3}.
\eemy
The coefficient $B$ is related to instanton effects. We 
find \cite{Schafer:2002ty}
\bemy
\label{B}
 B = c
 \left[\frac{3\sqrt{2}\pi}{g}\Delta
     \left(\frac{\mu^2}{2\pi^2}\right)\right]^2
  \left(\frac{8\pi^2}{g^2}\right)^{6}
  \frac{\Lambda_{QCD}^9}{\mu^{12}} ,
\eemy
which shows that $B$ is strongly suppressed at large chemical 
potential. The $A_i$ terms receive perturbative contributions
and are given by \cite{Son:1999cm,Schafer:2001za}
\bemy
 A_1= -A_2 = \frac{3\Delta^2}{4\pi^2}, 
\hspace{1cm} A_3 = 0.
\eemy
Finally, one can show that $X_L=MM^\dagger/(2p_F)$ and $X_R=M^\dagger M
/(2p_F)$ act as effective chemical potentials for left and right-handed 
fermion. These effective chemical potentials appear in the time derivative
of the chiral field \cite{Bedaque:2001je}, 
\bemy
\label{mueff}
 \nabla_0\Sigma = \partial_0 \Sigma 
 + i \left(\frac{M M^\dagger}{2p_F}\right)\Sigma
 - i \Sigma\left(\frac{ M^\dagger M}{2p_F}\right) .
\eemy

\section{Matter at non-asymptotic density}

 At non-asymptotic densities we can not rely on perturbative QCD 
calculations to determine the magnitude of the gap parameter, but 
the argument that pairing occurs and that the CFL phase is 
energetically favored is quite general, and does not depend on 
details of the interaction. The dominant stress on the CFL phase 
at non-asymptotic densities arises from flavor symmetry breaking 
due to the quark masses. We will focus on the physically relevant 
case $m_s\gg m_d\simeq m_u$. In this case the main expansion parameter 
is $m_s^2/(\mu\Delta)$, which is the ratio of the mass correction
to the Fermi energy of the strange quark over the magnitude of the gap.  

\begin{figure}[t]
\bc\includegraphics[width=7cm]{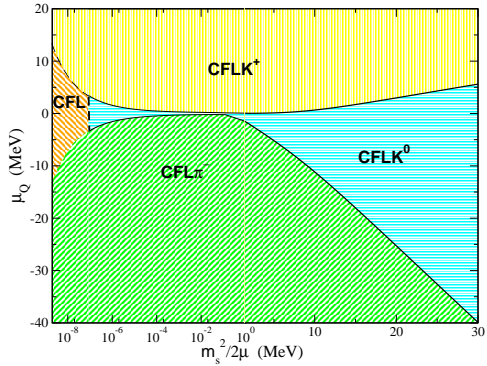}\ec
\caption{\label{fig_kcond}
Phase structure of CFL matter as a function of the effective 
chemical potential $\mu_s=m_s^2/(2p_F)$ and the lepton chemical 
potential $\mu_Q$, from Kaplan and Reddy \cite{Kaplan:2001qk}. 
A typical value of $\mu_s$ in a neutron star is 10 MeV. }
\end{figure}

\subsection{Kaon condensation}
\label{sec_kcond}

 Using the chiral effective lagrangian we can determine the 
dependence of the order parameter on the quark masses. The 
effective potential for the order parameter is 
\bemy
\label{v_eff}
V_{eff} = \frac{f_\pi^2}{4} {\rm Tr}\left[
 2X_L\Sigma X_R\Sigma^\dagger -X_L^2-X_R^2\right] 
     - A_1\left[ \left({\rm Tr}(M\Sigma^\dagger)\right)^2 
     - {\rm Tr}\left((M\Sigma^\dagger)^2\right) \right].
\eemy
The first term contains the effective chemical potential 
$\mu_s=m_s^2/(2p_F)$ and favors states with a deficit of 
strange quarks (with strangeness $S=-1$). The second term favors 
the neutral ground state $\Sigma=1$. The lightest excitation
with positive strangeness is the $K^0$ meson. We therefore
consider the ansatz $\Sigma = \exp(i\alpha\lambda_4)$ which
allows the order parameter to rotate in the $K^0$ direction. 
The vacuum energy is 
\bemy 
\label{k0+_V}
 V(\alpha) = -f_\pi^2 \left( \frac{1}{2}\left(\frac{m_s^2-m^2}{2p_F}
   \right)^2\sin(\alpha)^2 + (m_{K}^0)^2(\cos(\alpha)-1)
   \right),
\eemy
where $(m_K^0)^2= (4A_1/f_\pi^2)m(m+m_s)$. Minimizing the vacuum 
energy we obtain 
\bemy 
\cos(\alpha)= \left\{ \begin{array}{cl}
 1 & \mu_s<m_K^0 \\
\;\frac{(m_K^0)^2}{\mu_s^2}\; & \mu_s >m_K^0\\
\end{array}\right.
\eemy
Using the perturbative result for $A_1$ we can get an estimate of 
the critical strange quark mass. We find  
\bemy
\label{ms_crit}
 m_s (crit)= 3.03\cdot  m_d^{1/3}\Delta^{2/3},
\eemy
from which we obtain $m_s(crit)\simeq 70$ MeV for $\Delta\simeq 50$ 
MeV. This result suggests that strange quark matter at densities that 
can be achieved in neutron stars is kaon condensed. The phase structure 
as a function of the strange quark mass and non-zero lepton chemical 
potentials was studied by Kaplan and Reddy \cite{Kaplan:2001qk}, see 
Fig.~\ref{fig_kcond}. We observe that if the lepton chemical potential 
is non-zero charged kaon and pion condensates are also possible.

\subsection{Fermions in the CFL phase}
\label{sec_gCFL}

 Kaon condensation occurs for $\mu_s/\Delta\sim \sqrt{mm_s}/\mu
\ll 1$. For conditions relevant to neutron stars $\mu_s/\Delta$ 
can get significantly larger, reaching $\mu_s/\Delta \sim 1$. 
In this case some of the fermion modes may become gapless or 
almost gapless \cite{Alford:2003fq}. In order to study this regime
we have to include fermions in the effective field theory. The 
effective lagrangian for fermions in the CFL phase
is \cite{Kryjevski:2004jw,Kryjevski:2004kt}
\bea 
\label{l_bar}
{\mathcal L} &=&  
 {\rm Tr}\left(N^\dagger iv^\mu D_\mu N\right) 
 - D{\rm Tr} \left(N^\dagger v^\mu\gamma_5 
               \left\{ {\mathcal A}_\mu,N\right\}\right)
 - F{\rm Tr} \left(N^\dagger v^\mu\gamma_5 
               \left[ {\mathcal A}_\mu,N\right]\right)
  \nonumber \\
 & &  \mbox{} + \frac{\Delta}{2} \left\{ 
     \left( {\rm Tr}\left(N_LN_L \right) 
   - \left[ {\rm Tr}\left(N_L\right)\right]^2 \right)  
   - (L\leftrightarrow R) + h.c.  \right\}.
\eea
$N_{L,R}$ are left and right handed baryon fields in the 
adjoint representation of flavor $SU(3)$. The baryon fields 
originate from quark-hadron complementarity \cite{Schafer:1998ef}. 
We can think of $N$ as describing a quark which is surrounded 
by a diquark cloud, $N_L \sim q_L\langle q_L q_L\rangle$. The 
covariant derivative of the nucleon field is given by $D_\mu N
=\partial_\mu N +i[{\mathcal V}_\mu,N]$. The vector and axial-vector 
currents are 
\bemy
\label{v-av}
 {\mathcal V}_\mu = -\frac{i}{2}\left\{ 
  \xi \partial_\mu\xi^\dagger +  \xi^\dagger \partial_\mu \xi 
  \right\}, \hspace{1cm}
{\mathcal A}_\mu = -\frac{i}{2} \xi\left(\nabla_\mu 
    \Sigma^\dagger\right) \xi , 
\eemy
where $\xi$ is defined by $\xi^2=\Sigma$. $F$ and $D$ are low energy 
constants that determine the baryon axial coupling. In perturbative QCD 
we find $D=F=1/2$. The effective chemical potentials $(X_L,X_R)$ appear 
in the covariant derivative of the nucleon field. We have 
\bea
\label{V_X}
 D_0N     &=& \partial_0 N+i[\Gamma_0,N], \\
 \Gamma_0 &=& -\frac{i}{2}\left\{ 
  \xi \left(\partial_0+ iX_R\right)\xi^\dagger + 
  \xi^\dagger \left(\partial_0+iX_L\right) \xi 
  \right\}, \nonumber 
\eea
where $X_L=MM^\dagger/(2p_F)$ and $X_R=M^\dagger M/(2p_F)$ as before.
$(X_L,X_R)$ covariant derivatives also appears in the axial vector 
current given in equ.~(\ref{v-av}).

\begin{figure}[t]
\bc\includegraphics[width=9.5cm]{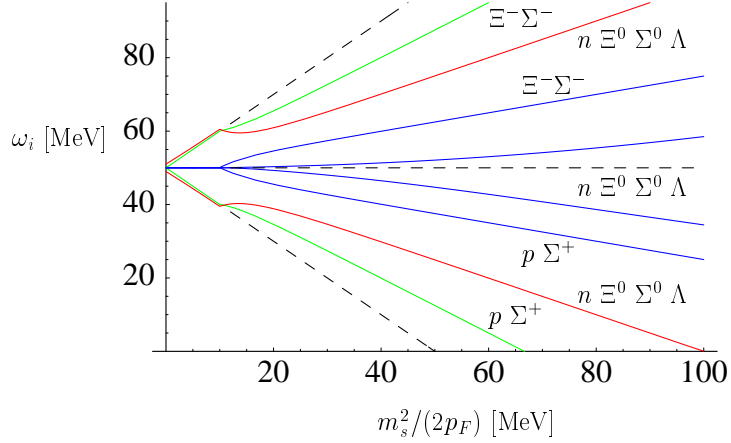}\ec
\caption{\label{fig_cfl_spec}
This figure shows the fermion spectrum in the CFL phase. For 
$m_s=0$ there are eight fermions with gap $\Delta$ and one
fermion with gap $2\Delta$ (not shown). Without kaon condensation
gapless fermion modes appear at $\mu_s=\Delta$ (dashed lines).
With kaon condensation gapless modes appear at $\mu_s=4\Delta/3$.}
\end{figure}

 We can now study how the fermion spectrum depends on the quark mass.
Since the field $N$ has the quark numbers of the baryon octet and 
singlet we will use $(p,n,\Sigma,\Xi,\Lambda)$ to label the fields.
In the CFL state we have $\xi=1$. For $\mu_s=0$ the octet has an 
energy gap $\Delta$ and the singlet has gap $2\Delta$. As a function 
of $\mu_s$ the excitation energy of the proton and neutron 
is lowered, $\omega_{p,n}=\Delta-\mu_s$, while the energy of the 
cascade states $\Xi^-,\Xi^0$ particles is raised, $\omega_{\Xi}=
\Delta+\mu_s$. All other excitation energies are independent of 
$\mu_s$. As a consequence we find gapless $(p,n)$ and $(\Xi^-,
\Xi^0)^{-1}$ excitations at $\mu_s=\Delta$. The situation is more 
complicated when kaon condensation is taken into account. In the kaon 
condensed phase there is mixing in the $(p,\Sigma^+,\Sigma^-,\Xi^-)$ 
and $(n,\Sigma^0,\Xi^0,\Lambda^8,\Lambda^0)$ sector. For $m_K^0\ll
\mu_s\ll \Delta$ the spectrum is 
given by
\bemy
\omega_{p\Sigma^\pm\Xi^-}= \left\{
 \begin{array}{c}
 \Delta \pm \frac{3}{4}\mu_s, \\
 \Delta \pm \frac{1}{4}\mu_s,
\end{array}\right.  \hspace{0.75cm}
\omega_{n\Sigma^0\Xi^0\Lambda} = \left\{
 \begin{array}{c}
   \Delta \pm \frac{1}{2}\mu_s ,\\ 
   \Delta , \\
   2\Delta .
 \end{array} \right. 
\eemy
Numerical results for the eigenvalues are shown in Fig.~\ref{fig_cfl_spec}. 
We observe that mixing within the charged and neutral baryon sectors leads 
to level repulsion. There are two modes that become light in the CFL window 
$\mu_s\leq 2\Delta$. One mode is a linear combination of proton and 
$\Sigma^+$ particles, as well as $\Xi^-$ and $\Sigma^-$ holes, and 
the other mode is a linear combination of the neutral baryons $(n,
\Sigma^0,\Xi^0,\Lambda^8,\Lambda^0)$.

\begin{figure}[t]
\bc\includegraphics[width=7.25cm]{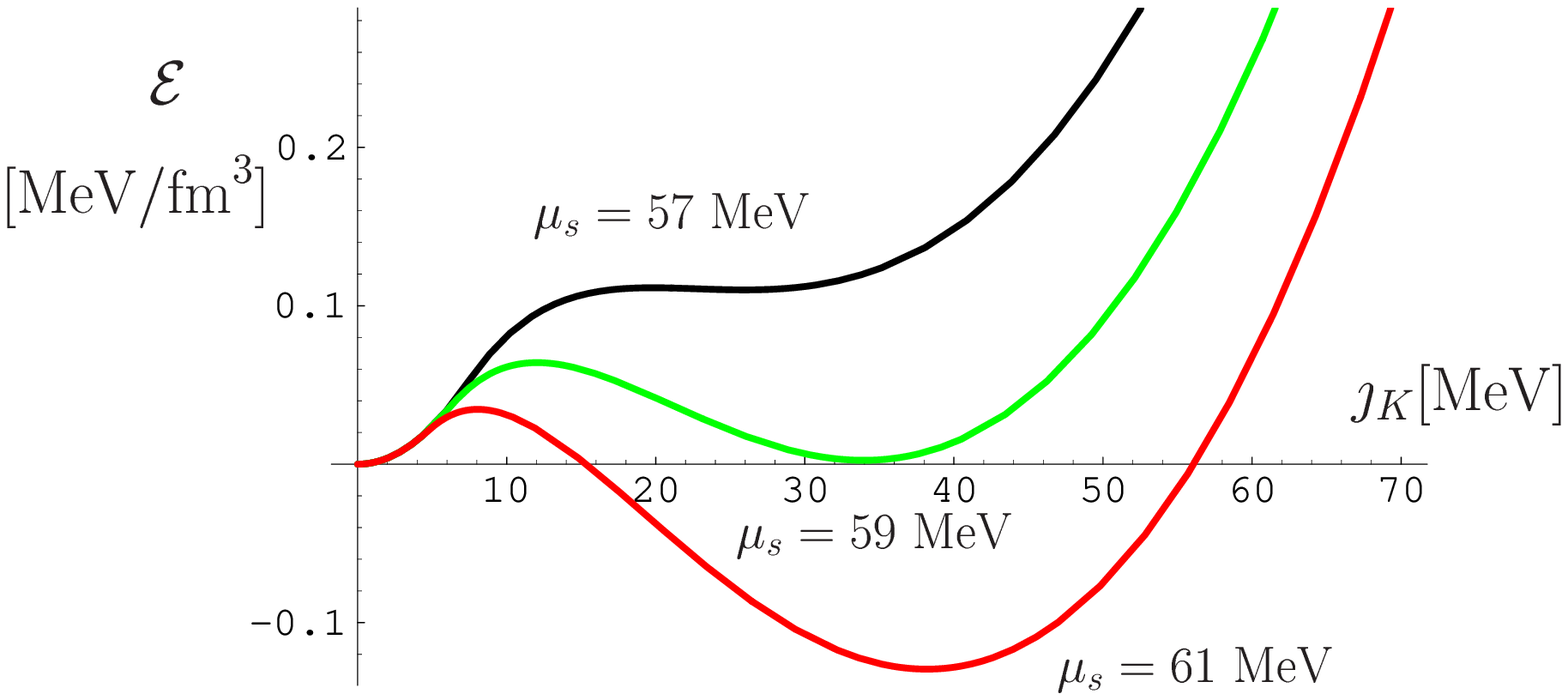}\hspace*{-0.3cm}
\includegraphics[width=7.25cm]{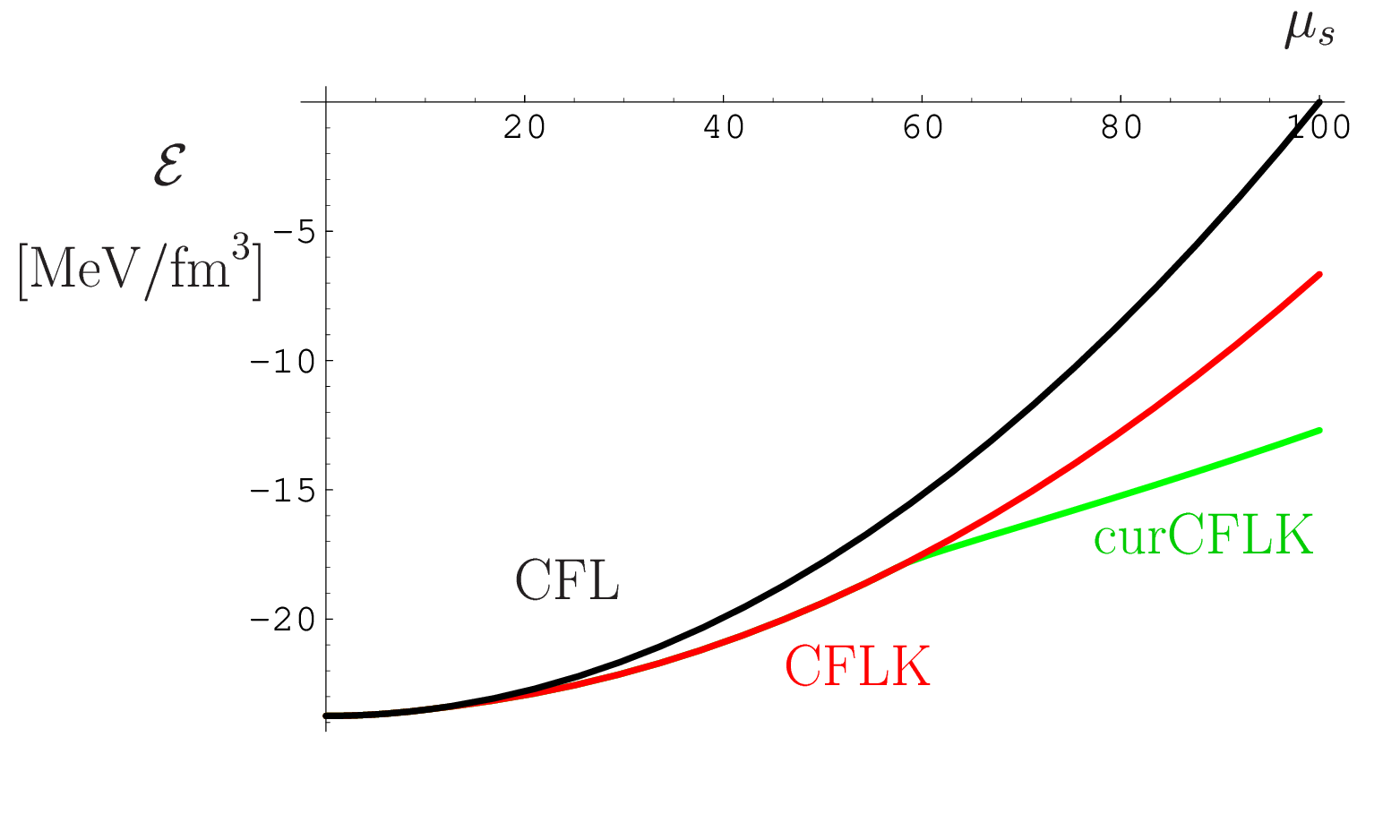}\ec
\caption{Left panel: Energy density as a function of the current 
$\jmath_K$ for several different values of $\mu_s=m_s^2/(2p_F)$ 
close to the phase transition. Right panel: Ground state energy 
density as a function of $\mu_s$. We show the CFL phase, the 
kaon condensed CFL (CFLK) phase, and the supercurrent state
(curCFLK).}
\label{fig_jfct}
\end{figure}

\subsection{Meson supercurrent state}
\label{sec_cur}

  What happens when gapless fermions appear in the spectrum? Several 
authors have shown that gapless fermion modes lead to instabilities 
in the current-current correlation function 
\cite{Huang:2004bg,Casalbuoni:2004tb}. Motivated by these results we 
have examined the stability of the kaon condensed phase against the 
formation of a non-zero current \cite{Schafer:2005ym,Kryjevski:2005qq}.
Consider a spatially varying $U(1)_Y$ rotation of the kaon condensate
\bemy 
U(x)\xi_{K^0} U^\dagger (x) = \left(
 \begin{array}{ccc}
 1 & 0 & 0 \\
 0 & 1/\sqrt{2} & ie^{i\phi_K(x)}/\sqrt{2} \\
 0 & ie^{-i\phi_K(x)}/\sqrt{2} & 1/\sqrt{2} 
\end{array} \right).
\eemy
This state is characterized by non-zero currents $\vec{\cal V}$ 
and $\vec{\cal A}$. In order to determine the stability of the 
CFLK state we compute the vacuum energy as a function of the kaon 
current $\vec{\jmath}_K=\vec\nabla\phi_K$. The meson part of the 
effective lagrangian gives a positive contribution
\bemy
{\cal E}=\frac{1}{2}v_\pi^2f_\pi^2\jmath_K^2 .
\eemy
A negative contribution can arise from gapless fermions. In order 
to determine this contribution we have to calculate the fermion spectrum 
in the presence of a non-zero current. The spectrum is determined 
by the effective lagrangian (\ref{l_bar}). The dispersion relation 
of the lowest mode is approximately given by
\bemy
\label{disp_ax}
\omega_l = \Delta +\frac{(l-l_0)^2}{2\Delta}-\frac{3}{4}
  \mu_s -\frac{1}{4}\vec{v}\cdot\vec{\jmath}_K,
\eemy
where $l=\vec{v}\cdot\vec{p}-p_F$ and we have expanded $\omega_l$ 
near its minimum $l_0=(\mu_s+\vec{v}\cdot\vec{\jmath}_K)/4$.
Equation (\ref{disp_ax}) shows that there is a gapless mode if 
$\mu_s>4\Delta/3-\jmath_K/3 $. The contribution of the gapless mode 
to the vacuum energy is 
\bemy
\label{e_fct} 
{\cal E} = \frac{\mu^2}{\pi^2}\int dl \int 
 \frac{d\Omega}{4\pi} \;\omega_l \theta(-\omega_l) ,
\eemy
where $d\Omega$ is an integral over the Fermi surface. In 
Fig.~\ref{fig_jfct} we show the ground state energy as a function 
of the current. We observe that there is a first order transition
to a state with a non-zero current. The ground state energy as a 
function of $\mu_s$ is shown in Fig.~\ref{fig_jfct}, see 
\cite{Gerhold:2006np,Gerhold:2006dt} for more details. Once the 
current becomes large the effective theory ceases to be reliable and 
states with multiple currents may appear. These states can be 
thought of as continuously connected to the crystalline quark 
matter phase \cite{Alford:2000ze}. 

\section{Transport Properties}
\label{sec_non_eq}

 Non-equilibrium properties, such as shear viscosity, bulk viscosity,
thermal conductivity and neutrino emissivity, play an important role 
in constraining the structure of compact stars. Shear and bulk 
viscosity control r-mode instabilities which, if not suppressed
by viscous damping, lead to a fast spin-down of rapidly rotating 
compact stars. Neutrino emissivity controls the cooling behavior 
of the star, and neutrino opacities determine the spectral shape of 
the initial neutrino burst. In addition to these specific constraints 
there has been significant progress, in both theory and observation,
of tying together the rotation of the star, the magnetic field, and
thermal properties. Exploiting these connections in order to 
constrain the phase structure of compact star matter will require 
a detailed understanding of transport properties.

\subsection{Hydrodynamics of the CFL phase}
\label{sec_hydro}

 The spontaneous breaking of $U(1)_B$ is related to superfluidity, 
and the $U(1)_B$ effective theory can be interpreted as superfluid 
hydrodynamics \cite{Son:2002zn}. We can define the fluid velocity as 
\bemy 
\label{def_v}
 v_\alpha = -\frac{1}{\mu_0} \, D_\alpha\varphi, 
\eemy
where $D_\alpha\varphi\equiv \partial_\alpha\varphi+(\mu,0)$
and $\mu_0\equiv (D_\alpha\varphi D^\alpha\varphi)^{1/2}$. Note 
that this definition ensures that the flow is irrotational, 
$\vec{\nabla}\times\vec{v}=0$. The identification (\ref{def_v}) is
motivated by the fact that the equation of motion for the $U(1)$
field $\varphi$ can be written as a continuity equation
\bemy
 \partial^\alpha  (n_0 v_\alpha) = 0,
\eemy
where $n_0 =3\mu_0^3/\pi^2$ is the superfluid number density. At 
$T=0$ the superfluid density is equal to the total density of 
the system, $n=dP/d\mu|_{\mu=\mu_0}$. The energy-momentum tensor
has the ideal fluid form 
\bemy
 T_{\alpha\beta} = (\epsilon+P)v_\alpha v_\beta -Pg_{\alpha\beta},
\eemy
and the conservation law $\partial^\alpha T_{\alpha\beta}=0$
corresponds to the relativistic Euler equation of ideal 
fluid dynamics. We conclude that the effective theory for the 
$U(1)_B$ Goldstone mode accounts for the defining characteristics 
of a superfluid: irrotational, non-dissipative hydrodynamic
flow. 

 At non-zero temperature the hydrodynamic description of a 
superfluid contains dissipative terms. Similar to the two fluid 
model of liquid helium we can describe the CFL phase, or any other 
relativistic superfluid, as a mixture of an ideal superfluid and a 
dissipative normal component \cite{Andersson:2001bz,Gusakov:2007px}. 
We will denote the densities of the superfluid and normal components 
by $\rho_s$ and $\rho_n$. We also define $u_\mu$ to be the velocity 
of the normal component, and $w_\mu$ to be the difference between 
the superfluid and normal velocities. The normal fluid provides
both non-dissipative and dissipative contributions to the 
energy momentum tensor. In the rest frame of the normal fluid 
the dissipative terms are 
\bemy
 \delta T_{ij} = -\eta \left( \partial_i u_j +\partial_j u_i
  - \frac{2}{3}\delta_{ij}\partial_k u_k \right) 
  - \delta_{ij} \zeta_1 \partial_k \left(\rho_s w_k\right)
  - \delta_{ij} \zeta_2 \partial_k u_k  \, 
\eemy
and
\bemy 
 \delta T_{0i} = -\kappa \partial_i T\ . 
\eemy
Here $\eta$ is the shear viscosity, $\zeta_{1,2}$ are bulk viscosities,
and $\kappa$ is the thermal conductivity. Two additional bulk 
viscosities, $\zeta_{3,4}$, control dissipative corrections to 
the equation of motion for the superfluid velocity. There is 
a symmetry relation between the kinetic coefficients that 
requires that $\zeta_4=\zeta_1$. Note that in the normal phase 
there is only one bulk viscosity, $\zeta\equiv\zeta_2$. 

\begin{figure}[t]
\begin{center} 
\includegraphics[width=7cm]{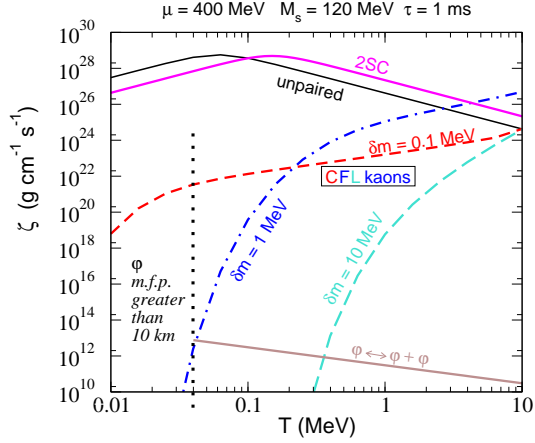}
\end{center}
\caption{Bulk viscosity $\zeta\equiv\zeta_2$ as a function of temperature 
for an oscillation period $\tau=2\pi/\omega=1$ ms. CFL phase: contribution
from the process $K^0\leftrightarrow \varphi+\varphi$ for different 
values of $\delta m\equiv m_{K^0}-\mu_s$ and contribution from 
$\varphi\leftrightarrow\varphi+\varphi$. 2SC phase and unpaired quark 
matter: contribution from the process $u+d\leftrightarrow u+s$.}
\label{fig:bulkCFL}
\end{figure}

\subsection{Transport Coefficients}
\label{sec_trans}

 The normal fluid is composed of quasi-particle excitations. 
In the CFL phase all quark modes are gapped and the relevant 
excitations are Goldstone bosons. At very low temperature, transport 
properties are dominated by the massless Goldstone boson $\varphi$ 
associated with the breaking of the $U(1)_B$ symmetry. The effective 
lagrangian (\ref{u1_eft_3}) determines the rates for the relevant 
scattering processes. 

 Shear viscosity is related to momentum transport. At low temperature
the shear viscosity of the CFL phase is determined by  $\varphi+\varphi 
\leftrightarrow\varphi+\varphi$ scattering. Manuel et al.~find 
\cite{Manuel:2004iv} 
\bemy
\label{eta_cfl}
\eta \simeq 1.3\times 10^{-4} \, \frac{\mu^8}{T^5} \, . 
\eemy
The bulk viscosity is sensitive to particle number changing processes.
This includes purely strong decays like $\varphi\leftrightarrow
\varphi+\varphi$, or electroweak processes like the strangeness
changing reaction $K^0\to\varphi+\varphi$. We first consider 
the pure QCD contribution. Bulk viscosity vanishes in an exactly 
scale invariant system. For realistic quark masses the dominant 
source of scale breaking is the strange quark mass. The contribution 
from the process $\varphi\leftrightarrow \varphi+\varphi$ is 
\cite{Manuel:2007pz}
\bemy
\label{zeta_cfl}
\zeta_2 \simeq 0.011 \frac{m_s^4}{T}  \, .
\eemy
We show this contribution in Fig.~\ref{fig:bulkCFL}. The electroweak
process $K^0\leftrightarrow \varphi+\varphi$ was studied in 
\cite{Alford:2007rw}. The weak contribution has a significant 
frequency dependence. In Fig.~\ref{fig:bulkCFL} we show the results
for an oscillation period $\tau=2\pi/\omega=1$ ms. We observe that 
at $T\simeq (1-10)$ MeV the bulk viscosity of CFL matter is comparable 
to that of unpaired quark matter. For $T<1$~MeV, $\zeta_2$ is strongly 
suppressed. Depending on the poorly known value for $\delta m\equiv 
m_{K^0}-\mu_s$ the pure $\varphi$ contribution given in 
equ.~(\ref{zeta_cfl}) may dominate over the $K^0\leftrightarrow 
\varphi+\varphi$ reaction at low enough temperatures. However, for 
$T<0.1$ MeV the $\varphi$ mean free path is on the order of the size 
of the star, i.e., the system is in the collisionless rather than 
in the hydrodynamic regime, and the result ceases to be meaningful. 

 The thermal conductivity of a CFL superfluid was studied by Braby 
et al.~\cite{Braby:2009dw}. The calculation is subtle because 
$\kappa$ vanishes for a system of quasi-particles with exactly 
linear dispersion relations \cite{Khalatnikov:1965}. The reason 
is that $\kappa$ measures the rate of energy transport relative 
to the motion of the fluid, but in a gas of massless particles with 
linear dispersion one cannot transport energy without transporting 
momentum. As a consequence, thermal conductivity is sensitive
to non-linearities in the dispersion relation. Braby et al.~find
\cite{Braby:2009dw}
\bemy
\kappa \simeq  4.01 \times 10^{-2}\, 
  \frac{\mu^8}{\Delta^6}\,{\rm MeV}^2\, . 
\eemy
They also estimate the contribution to $\kappa$ from phonon scattering 
on kaons. This term grows as $\sqrt{T}$, but it is significantly 
smaller than the phonon contribution in the regime where the calculation 
is reliable. Non-linearities in the dispersion relation also play a 
role in determining the remaining two bulk viscosities, $\zeta_1$ and 
$\zeta_3$ \cite{Mannarelli:2009ia}. Mannarelli and Manuel find 
$\zeta_1\sim m_s^2/(\mu T)$ and $\zeta_3\sim 1/(T\mu^2)$. Note 
that $\zeta_3$ is non-zero even in the approximately conformal 
limit $m_s\to 0$. 

\subsection{Neutrino emissivity}
\label{sec_emis}

In CFL quark matter the neutrino emissivity is dominated by 
reactions involving pseudo-Goldstone modes such as
\bea
\pi^\pm,K^\pm &\to& e^\pm +\bar{\nu}_e \, , \nonumber \\
\pi^0 &\to& \nu_e +\bar{\nu}_e \, , \\
\varphi + \varphi &\to& \varphi + \nu_e +\bar{\nu}_e \, . \nonumber
\eea
These processes were studied in \cite{Jaikumar:2002vg,Reddy:2002xc}. 
The decay rates of the massive mesons $\pi^\pm$, $K^\pm$, and $\pi^0$ 
are proportional to their number densities and are suppressed by
Boltzmann factors $\exp(-E/T)$, where $E$ is the energy gap of the 
meson. The emissivity from $\pi^\pm$ decay is
\bemy
  \epsilon_\pi = \frac{1}{8\pi}\,
    (G_F^2\, f_{\pi}^2\, m_e^2)\, m_\pi^2\, n_{\pi}\,
    \left( 1 + 2\left(1-v_\pi^2\right)
      + \frac{2m_\pi T}{v_{\pi}^2m_e^2} \left(1-v_\pi^2\right)^2
   \right)\,,
\label{emisspiappx}
\eemy
where $n_\pi$ is the number density of pions. Similar results can be
derived for $\pi^0$ and $K^\pm$ decay. Since the pseudo-Goldstone 
boson energy gaps are on the order of a few MeV, the emissivities are 
strongly suppressed as compared to unpaired quark matter for temperatures 
below this scale. Neutrino emission from processes involving the 
$\varphi$ is not exponentially suppressed, but it involves a very 
large power of $T$, 
\bemy
\epsilon_\nu \sim  \frac{G_F^2 T^{15}}{f^2\mu^4}\, , 
\eemy
and is numerically very small. Reddy et al.~also studied the neutrino 
mean free path $l_\nu$. For $T\sim 30$ MeV the mean free path is on 
the order of 1 m, but for $T<1$~MeV, $l_\nu>10$ km \cite{Reddy:2002xc}.

\section{Outlook}
\label{sec_out}

 There are a variety of issues that remain to be studied. While 
the calculation of transport coefficients in the CFL phase is 
now essentially complete, this is not the case for many of the 
less dense phases. There are calculations of shear and bulk 
viscosity as well as neutrino emissivity in the CFL-K phase
\cite{Reddy:2003ap,Alford:2009jm,Alford:2008pb}, but there 
are essentially no results for the spatially inhomogeneous or 
anisotropic phases. There is also much work to be done in 
order to understand many phenomena that are relevant in 
compact stars, like mutual friction between the normal fluid
and superfluid vortices \cite{Mannarelli:2008je,Haskell:2009fz}, 
or the role of the quark-hadron interface. 

Acknowledgments: This work was carried out in collaboration with 
M.~Alford, P.~Bedaque, M.~Braby, J.~Chao, A.~Gerhold, A.~Kryjevski, 
and S.~Reddy. The work was supported in part by the US Department 
of Energy grant DE-FG02-03ER41260.


\begin{thebibliography}{99}



\bibitem{Alford:1999mk}
M.~Alford, K.~Rajagopal and F.~Wilczek,
Nucl.\ Phys.\  {\bf B537}, 443 (1999)
[hep-ph/9804403].

\bibitem{Bedaque:2001je}
P.~F.~Bedaque and T.~Sch{\"a}fer,
Nucl.\ Phys.\ {\bf A697}, 802 (2002)
[hep-ph/0105150].

\bibitem{Schafer:2005ym}
T.~Sch{\"a}fer,
Phys.\ Rev.\ Lett.\ {\bf 96}, 012305 (2006)
[hep-ph/0508190].

\bibitem{Kryjevski:2005qq}
A.~Kryjevski,
Phys.\ Rev.\  D {\bf 77}, 014018 (2008)
[arXiv:hep-ph/0508180].

\bibitem{Alford:2007xm}
M.~G.~Alford, A.~Schmitt, K.~Rajagopal and T.~Sch\"afer,
Rev.\ Mod.\ Phys.\  {\bf 80}, 1455 (2008)
[arXiv:0709.4635 [hep-ph]].

\bibitem{Alford:2000ze}
M.~G.~Alford, J.~A.~Bowers and K.~Rajagopal,
Phys.\ Rev.\ D {\bf 63}, 074016 (2001)
[hep-ph/0008208].

\bibitem{Casalbuoni:2003wh}
R.~Casalbuoni and G.~Nardulli,
Rev.\ Mod.\ Phys.\  {\bf 76}, 263 (2004)
[hep-ph/0305069].

\bibitem{Schafer:2000tw}
T.~Sch\"afer,
Phys.\ Rev.\  D {\bf 62}, 094007 (2000)
[arXiv:hep-ph/0006034].

\bibitem{Schmitt:2004et}
A.~Schmitt,
Phys.\ Rev.\  D {\bf 71}, 054016 (2005)
[arXiv:nucl-th/0412033].

\bibitem{Rapp:1997zu}
R.~Rapp, T.~Sch\"afer, E.~V.~Shuryak and M.~Velkovsky,
Phys.\ Rev.\ Lett.\  {\bf 81}, 53 (1998)
[arXiv:hep-ph/9711396].

\bibitem{Alford:1997zt}
M.~G.~Alford, K.~Rajagopal and F.~Wilczek,
Phys.\ Lett.\  B {\bf 422}, 247 (1998)
[arXiv:hep-ph/9711395].

\bibitem{Deryagin:1992rw}
D.~V.~Deryagin, D.~Y.~Grigoriev and V.~A.~Rubakov,
Int.\ J.\ Mod.\ Phys.\  A {\bf 7}, 659 (1992).

\bibitem{Shuster:1999tn}
E.~Shuster and D.~T.~Son,
Nucl.\ Phys.\  B {\bf 573}, 434 (2000)
[arXiv:hep-ph/9905448].

\bibitem{Kojo:2009ha}
T.~Kojo, Y.~Hidaka, L.~McLerran and R.~D.~Pisarski,
arXiv:0912.3800 [hep-ph].

\bibitem{Shovkovy:2003uu}
I.~Shovkovy and M.~Huang,
Phys.\ Lett.\  B {\bf 564}, 205 (2003)
[arXiv:hep-ph/0302142].

\bibitem{Alford:2003fq}
M.~Alford, C.~Kouvaris and K.~Rajagopal,
Phys.\ Rev.\ Lett.\  {\bf 92}, 222001 (2004)
[hep-ph/0311286].

\bibitem{Alford:2004pf}
M.~Alford, M.~Braby, M.~W.~Paris and S.~Reddy,
Astrophys.\ J.\  {\bf 629}, 969 (2005)
[arXiv:nucl-th/0411016].

\bibitem{Kurkela:2009gj}
A.~Kurkela, P.~Romatschke and A.~Vuorinen,
arXiv:0912.1856 [hep-ph].

\bibitem{Lattimer:2006xb}
J.~M.~Lattimer and M.~Prakash,
Phys.\ Rept.\  {\bf 442}, 109 (2007)
[arXiv:astro-ph/0612440].

\bibitem{Danielewicz:2002pu}
P.~Danielewicz, R.~Lacey and W.~G.~Lynch,
Science {\bf 298}, 1592 (2002)
[arXiv:nucl-th/0208016].

\bibitem{Schafer:1999fe}
T.~Sch{\"a}fer,
Nucl.\ Phys.\ B {\bf 575}, 269 (2000)
[hep-ph/9909574].

\bibitem{Evans:1999at}
N.~Evans, J.~Hormuzdiar, S.~D.~Hsu and M.~Schwetz,
Nucl.\ Phys.\ B {\bf 581}, 391 (2000)
[hep-ph/9910313].

\bibitem{Son:1999uk}
D.~T.~Son,
Phys.\ Rev.\  {\bf D59}, 094019 (1999)
[hep-ph/9812287].
 
\bibitem{Brown:1999aq}
W.~E.~Brown, J.~T.~Liu and H.~C.~Ren,
Phys.\ Rev.\ D {\bf 61}, 114012 (2000)
[hep-ph/9908248].

\bibitem{Wang:2001aq}
Q.~Wang and D.~H.~Rischke,
Phys.\ Rev.\ D {\bf 65}, 054005 (2002)
[nucl-th/0110016].

\bibitem{Schafer:2003jn}
T.~Sch{\"a}fer,
Nucl.\ Phys.\ A {\bf 728}, 251 (2003)
[hep-ph/0307074].
 
\bibitem{Schafer:1999jg}
T.~Sch{\"a}fer and F.~Wilczek,
Phys.\ Rev.\  {\bf D60}, 114033 (1999)
[hep-ph/9906512].
  
\bibitem{Buballa:2003qv}
M.~Buballa,
Phys.\ Rept.\  {\bf 407}, 205 (2005)
[arXiv:hep-ph/0402234].

\bibitem{Son:1999cm}
D.~T.~Son and M.~Stephanov, 
Phys.\ Rev.\ {\bf D61}, 074012 (2000) 
[hep-ph/9910491], 
erratum: hep-ph/0004095.

\bibitem{Rischke:2000ra}
D.~H.~Rischke,
Phys.\ Rev.\  D {\bf 62}, 054017 (2000)
[arXiv:nucl-th/0003063].

\bibitem{Manuel:2001mx}
C.~Manuel and K.~Rajagopal,
Phys.\ Rev.\ Lett.\  {\bf 88}, 042003 (2002)
[arXiv:hep-ph/0107211].

\bibitem{Son:2002zn}
  D.~T.~Son,
  preprint, hep-ph/0204199.

\bibitem{Casalbuoni:1999wu}
R.~Casalbuoni and D.~Gatto,
Phys.\ Lett.\ {\bf B464}, 111 (1999)
[hep-ph/9908227].

\bibitem{Schafer:2002ty} 
T.~Sch{\"a}fer,
Phys.\ Rev.\ D {\bf 65}, 094033 (2002)
[hep-ph/0201189].

\bibitem{Schafer:2001za}
T.~Sch\"afer,
Phys.\ Rev.\ D {\bf 65}, 074006 (2002)
[hep-ph/0109052].

\bibitem{Kaplan:2001qk}
D.~B.~Kaplan and S.~Reddy,
Phys.\ Rev.\ D {\bf 65}, 054042 (2002)
[hep-ph/0107265].

\bibitem{Kryjevski:2004jw}
A.~Kryjevski and T.~Sch{\"a}fer,
Phys.\ Lett.\ B {\bf 606}, 52 (2005)
[hep-ph/0407329].

\bibitem{Kryjevski:2004kt}
A.~Kryjevski and D.~Yamada,
Phys.\ Rev.\ D {\bf 71}, 014011 (2005)
[hep-ph/0407350].

\bibitem{Schafer:1998ef}
T.~Sch{\"a}fer and F.~Wilczek,
Phys.\ Rev.\ Lett.\  {\bf 82}, 3956 (1999)
[hep-ph/9811473].

\bibitem{Huang:2004bg}
M.~Huang and I.~A.~Shovkovy,
Phys.\ Rev.\ D {\bf 70}, 051501 (2004)
[hep-ph/0407049].

\bibitem{Casalbuoni:2004tb}
R.~Casalbuoni, R.~Gatto, M.~Mannarelli, G.~Nardulli and M.~Ruggieri,
Phys.\ Lett.\ B {\bf 605}, 362 (2005)
[hep-ph/0410401].

\bibitem{Gerhold:2006np}
A.~Gerhold, T.~Sch\"afer and A.~Kryjevski,
Phys.\ Rev.\  D {\bf 75}, 054012 (2007)
[arXiv:hep-ph/0612181].

\bibitem{Gerhold:2006dt}
A.~Gerhold and T.~Sch\"afer,
Phys.\ Rev.\  D {\bf 73}, 125022 (2006)
[arXiv:hep-ph/0603257].

\bibitem{Andersson:2001bz}
N.~Andersson and G.~L.~Comer,
Mon.\ Not.\ Roy.\ Astron.\ Soc.\  {\bf 328}, 1129 (2001)
[arXiv:astro-ph/0101193].

\bibitem{Gusakov:2007px} 
M.~E.~Gusakov,
Phys.\ Rev.\  D {\bf 76}, 083001 (2007)
[arXiv:0704.1071 [astro-ph]].

\bibitem{Manuel:2004iv}
C.~Manuel, A.~Dobado and F.~J.~Llanes-Estrada,
JHEP {\bf 0509}, 076 (2005)
[arXiv:hep-ph/0406058].

\bibitem{Manuel:2007pz}
C.~Manuel and F.~J.~Llanes-Estrada,
JCAP {\bf 0708}, 001 (2007)
[arXiv:0705.3909 [hep-ph]].

\bibitem{Alford:2007rw}
M.~G.~Alford, M.~Braby, S.~Reddy and T.~Sch\"afer,
Phys.\ Rev.\  C {\bf 75}, 055209 (2007)
[arXiv:nucl-th/0701067].

\bibitem{Braby:2009dw}
M.~Braby, J.~Chao and T.~Sch\"afer,
arXiv:0909.4236 [hep-ph].

\bibitem{Khalatnikov:1965}
I.~M.~Khalatnikov,
``Introduction to the Theory of Superfluidity'',
W.~.A.~Benjamin, Inc. (1965). 

\bibitem{Mannarelli:2009ia}
M.~Mannarelli and C.~Manuel,
arXiv:0909.4486 [hep-ph].

\bibitem{Jaikumar:2002vg}
P.~Jaikumar, M.~Prakash and T.~Sch\"afer,
Phys.\ Rev.\  D {\bf 66}, 063003 (2002)
[arXiv:astro-ph/0203088].

\bibitem{Reddy:2002xc}
S.~Reddy, M.~Sadzikowski and M.~Tachibana,
Nucl.\ Phys.\  A {\bf 714}, 337 (2003)
[arXiv:nucl-th/0203011].

\bibitem{Reddy:2003ap}
S.~Reddy, M.~Sadzikowski and M.~Tachibana,
Phys.\ Rev.\  D {\bf 68}, 053010 (2003)
[arXiv:nucl-th/0306015].

\bibitem{Alford:2009jm}
M.~G.~Alford, M.~Braby and S.~Mahmoodifar,
arXiv:0910.2180 [nucl-th].

\bibitem{Alford:2008pb}
M.~G.~Alford, M.~Braby and A.~Schmitt,
J.\ Phys.\ G {\bf 35}, 115007 (2008)
[arXiv:0806.0285 [nucl-th]].

\bibitem{Mannarelli:2008je}
M.~Mannarelli, C.~Manuel and B.~A.~Sa'd,
Phys.\ Rev.\ Lett.\  {\bf 101}, 241101 (2008)
[arXiv:0807.3264 [hep-ph]].

\bibitem{Haskell:2009fz}
B.~Haskell, N.~Andersson and A.~Passamonti,
arXiv:0902.1149 [astro-ph.SR].

\end{thebibliography}
\end{document}